# Single-Shot Orientation Imaging of Nanorods Using Spin-to-Orbital Angular Momentum Conversion of Light

Tomáš Fordey,[1] Petr Bouchal,[2,3,*] Michal Baránek,[1] Petr Schovánek,[1] Zdeněk Bouchal,[1] Petr Dvořák,[2,3] Katarína Rovenská,[2,3] Filip Ligmajer,[2,3] Radim Chmelík,[2,3] Tomáš Šikola[2,3]

[1]*Department of Optics, Palacký University, 17. listopadu 1192/12, 771 46 Olomouc, Czech Republic*
[2]*Institute of Physical Engineering, Faculty of Mechanical Engineering, Brno University of Technology, Technická 2, 616 69 Brno, Czech Republic*
[3]*Central European Institute of Technology, Brno University of Technology, Purkyňova 656/123, 612 00 Brno, Czech Republic*

*\*petr.bouchal@ceitec.vutbr.cz*

## Abstract

The key information about any nanoscale system are orientations and conformations of its parts. Unfortunately, these details are often hidden below the diffraction limit and elaborate techniques must be used to optically probe them. Here, we present a single-shot imaging technique allowing time-resolved monitoring of rotation motion of metal nanorods, realized in a wide-field regime and with no ambiguity of the measured angles. In our novel method, the nanorod orientation is imprinted onto a geometric phase of scattered light composed of the opposite spin states. By spin-to-orbital angular momentum conversion, we generate two oppositely winding helical waves (optical vortices) that are used for restoring the nanorod in-plane orientation. The method was calibrated using lithographically fabricated nanorods and tested by the rotation imaging of immobilized and moving sub-100 nm colloidal nanorods (measurement accuracy of 2.5°). We envision this technique can be used also for estimation of nanorod aspect ratios and their out-of-plane orientations.

## Introduction

Nanoparticles fabricated from noble metals exhibit remarkable shape-dependent optical properties associated with the excitation of surface plasmon resonance (SPR).[1,2] Novel applications using nanoparticles as contrast agents, imaging and sensing probes have emerged thanks to their biocompatibility, ability to enhance the driving field without bleaching, and SPR sensitivity to the nanoparticle size and its surrounding environment.[3,4] In recent years, elongated nanoparticles (nanorods) exhibiting longitudinal SPR (LSPR) have attracted a growing interest. The nanorods became popular thanks to weak radiation damping due to small volumes, high light-scattering efficiencies and large local-field enhancement factors,[5] and the possibility to manipulate their aspect ratio.[6] Thanks to the shape anisotropy, the nanorods were deployed as sensitive orientation sensors in the conformation and dynamics studies of biological systems.[7] The orientation imaging goes beyond biophotonic applications and can be used as a powerful tool for the study of nanorod supercrystals,[8] superlattices,[9] or nanorod chains[10] deployed in new plasmonic-based nano-imaging devices.[11] In previous works, the problem of orientation imaging was addressed by different modifications of dark-field polarization microscopy,[12–17] polarization-sensitive photothermal imaging,[7] or confocal microscopy combined with higher-order laser modes.[18] In the dark-field microscopy methods, the orientation measurement was realized by comparing defocused images with a series of simulated images,[12] or using successive images recorded with different polarization directions of illumination,[13–15] otherwise the orientation ambiguity arose.[16] In the confocal microscopy approach to orientation imaging, scanning by a doughnut laser beam along the field of view was needed. However, the laser light and speckles accompanying its spatial coherence may be a barrier for applications using nanorods as biological probes. Significant progress in orientation imaging was achieved with differential interference contrast (DIC) microscopy.[19,20] Using DIC, the orientation imaging was combined with three-dimensional tracking, but the impossibility to distinguish the rotational motion of the nanorod from its axial movement required deploying parallax microscopy and a feedback loop algorithm to control the focal plane position.[21]

Since optical microscopy is limited by light diffraction, orientation imaging of nanorods is a powerful way allowing to optically enter the nanoscale and study biological[7,22,23] or artificial nanosystems[10,11,24] through orientation changes of deeply subdiffraction light emitters. The problem of orientation imaging can be further generalized to single-molecule fluorescence microscopy,[25–27] study of fluorescent nanoemitters[28] or dipole orientation in solid-state single photon emitters.[29] Real-world applications often rely on the study of systems undergoing dynamic changes. Therefore, the previously unavailable real-time and accurate orientation imaging without ambiguity of measured angles is of great demand. In this letter, we capitalize on the recent progress in (nano)photonics that has brought a new types of light fields with spatially structured amplitude,[30] phase,[31] or polarization,[32,33] while opening new possibilities for discovering sophisticated yet simple to implement imaging modalities. We present a new strategy of single-shot orientation imaging stemming from the control of spatially structured states of light. In our experiments, the spin angular momentum (SAM) of circularly polarized light is converted to the orbital angular momentum (OAM) associated with the wavefront helicity of optical vortices.[34–36] This is achieved through spatially varying anisotropy, which is affecting the geometric (Pancharatnam-Berry) phase in polarization-sensitive optical components prepared by fourth-generation technologies (so called 4G optics).[37,38] Using the geometric-phase optical element for SAM to OAM conversion,[39] often referred to as q-plate,[40] a pair of optical vortices is generated from LSPR light, whose subsequent interference pattern provides information on the nanorod angular orientation. The diffraction image spots formed by the vortex interference are created simultaneously for all nanorods in the field of view and their angular orientation is determined without ambiguity[16] from a single image acquired using dark-field illumination. This is a significant step forward compared to previously used techniques, opening new possibilities for studying very fast events such as conformational dynamics of biological cells,[22] rotation of biological motor proteins[23] or nanorod motors.[24] Thanks to the use of incoherent light, the developed technique is free of coherent artifacts and the evaluation of the rotational motion of the nanoparticle is not affected by its axial displacement. This is an important advantage over the DIC method, which signal does not allow distinguishing between these two kinds of movements. Optical performance of our method is demonstrated by a calibration experiment with lithographically fabricated gold nanorods and by orientational imaging of colloidal sub-100 nm gold nanorods (diameter 25 nm and length 71 nm). Using pre-defined orientations of the benchmark sample, the evaluated accuracy fell below 4.5°, while scanning electron microscopy images of colloidal nanorods proved the possibility of measurement with accuracy 2.5°.

**Results**

*Experimental strategy*

Thanks to its anisotropic shape, a nanorod can support two SPR modes corresponding to transverse SPR and LSPR, respectively. The resonant wavelengths of both modes are spectrally shifted and thus can be excited independently. The proposed orientation imaging method takes advantage of the dominant LSPR. The nanorod is considered as a dipolar light source that emits linearly polarized light with the electric field oscillating in the direction of the excited LSPR (Fig. 1). The linear polarization can be described by the state vector $|\varphi\rangle$, determined by the angle $\varphi$ of the polarization direction in the x-y plane. The linearly polarized light does not carry SAM but can be decomposed into left- and right-handed circular polarizations, $|L\rangle$ and $|R\rangle$, carrying opposite SAM given by $\pm\hbar$ per photon ($\hbar$ is the reduced Planck constant). When the light from a nanorod oriented at the angle $\varphi$ is expressed in the used vector basis, the phase shifts $\pm\varphi$ are imposed on $|L\rangle$ and $|R\rangle$ states.[37,38] This means that the geometric-phase difference $2\varphi$ is introduced between the circular polarizations, which is assigned to the $|L\rangle$ state in our convention (Fig. 1). By passing a q-plate,[40] the light emitted by the nanorod undergoes SAM to OAM conversion, in which the reversion of the circular polarizations is accompanied by generation of optical vortex beams. The vortex beams are characterized by helical wavefronts, meaning the light acts as twisted around the optical axis and each of its photons carries OAM of $l\hbar$. The integer $l$ is called the topological charge and specifies how many twists the wavefront makes per one wavelength. In our experiment, the light with polarization $|L\rangle$ ($+\hbar$ SAM) is transformed to a vortex beam with polarization $|R\rangle$ ($-\hbar$ SAM) and the helical wavefront with the topological charge $l = 1$ ($+\hbar$ OAM). A similar operation is simultaneously carried out with the $|R\rangle$ input state; hence a pair of vortices with $l = \pm 1$ is generated from the linearly polarized light emitted by individual nanorods (see Fig. 1). If recorded on a screen, the vortex beam would form a characteristic ring ($|A|^2$ in Fig. 1) because in its center a phase singularity occurs. Instead of directly recording these orthogonally polarized vortex beams, we project their electric fields into the same direction by a linear polarizer and let them interfere. That way, images of the individual nanorods are created with a characteristic azimuthal two-lobe pattern because the interfering helical wavefronts with opposite windings have an azimuthally varying phase difference. The geometric-phase difference, imprinted in the two vortex beams by the nanorod orientation, is thus translated into the angular orientation of the two-lobe optical image of a sub-diffraction point-like source (known in other contexts as a double-helix point spread function, DH PSF[41]). By evaluating the rotation of each nanorod's DH PSF, its instantaneous angular position can be easily determined. Implementation of this scheme into a modified dark-field microscope would thus enable real-time tracking of nanorod positions and orientations in the whole view field, a task which is currently unattainable by any other means. Note that in previous studies, the rotating DH PSF was used for measuring the axial position of point emitters.[41] However, the DH PSF generated in our system is independent of defocusing and its orientation is driven exclusively by the angular positioning of the nanorod. We will justify this claim in the following paragraph along with a detailed theoretical description of the measurement principle.

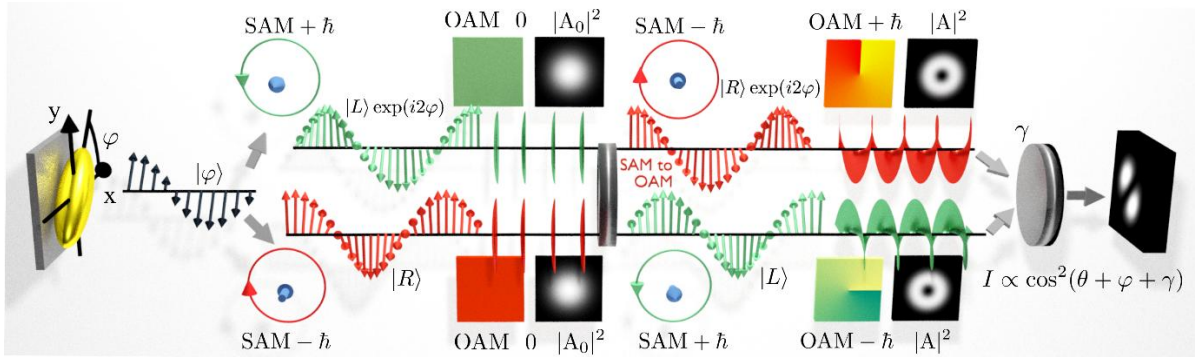

**Figure 1** Principle of orientation imaging using SAM to OAM conversion. Linearly polarized light $|\varphi\rangle$ from a nanorod (located in x-y plane at an angle φ) can be decomposed into $|L\rangle$ and $|R\rangle$ circular polarizations (SAM of $\pm\hbar$). Upon passing through a q-plate and undergoing a SAM to OAM conversion, their handedness is reversed to create co-propagating optical vortices with topological charges of $l = \pm 1$ (OAM of $\pm\hbar$). Finally, the two beams pass a polarizer, which enables their interference, and a two-lobe interference image spot (DH PSF) is formed that allows measuring the nanorod orientation.

The rotation of the DH PSF can be explained by a simplified mathematical model, considering an ideal dipole emitter free from background reflection of the exciting light. This assumption is satisfied in dark-field microscopy configuration if only a single SPR mode dominates (i.e., LSPR here). The scattered linearly polarized light, carrying information about the nanorod orientation angle $\varphi$ within the x-y plane, is described by a state vector $|\varphi\rangle = |x\rangle\cos\varphi + |y\rangle\sin\varphi$, normalized as $\langle\varphi|\varphi\rangle = 1$. When decomposed into the base of circular polarizations instead, $|\varphi\rangle = e^{-i\varphi}(|L\rangle e^{i2\varphi} + |R\rangle)/2$, a geometric-phase difference $2\varphi$ is introduced between the base vectors.[37,38] The light emitted from the nanorod is collected by the used optical system and transformed by a q-plate.[40] During this transformation, the input states $|L\rangle$ and $|R\rangle$ are swapped and two vortex beams emerge as the transmitted beams. These beams have characteristic helical wavefronts[34–36] with azimuthally changing phases $e^{i\theta}$ and $e^{-i\theta}$ attached to the transmitted $|R\rangle$ and $|L\rangle$ polarizations, respectively. A diffraction-limited image of the nanorod is thus created, represented by a state vector $|\varphi'\rangle = Ae^{-i\varphi}(|R\rangle e^{i(\theta+2\varphi)} + |L\rangle e^{-i\theta})/2$, where $|A|^2$ represents a vortex PSF of the imaging system with a typical doughnut-shaped intensity profile. Prior to the detection by the microscope camera, the orthogonal polarization states $|L\rangle$ and $|R\rangle$ are recombined using a linear polarizer so the two vortices can interfere. The diffraction-limited image of the nanorod then emerges as a sum of two vortex PSFs with intensity azimuthally modulated by the cross-correlation of vortex beams with opposite helical wavefronts: $I = 2|A|^2\cos^2(\theta + \varphi + \gamma)$, where $\gamma$ is the angle giving the transmission direction of the polarizer. Note the azimuthal orientation of the DH PSF remains unchanged under defocusing (the cosine argument is independent of a depth coordinate), and only some size scaling of the lobes can occur (scaling in $|A|^2$ available by analytical calculations). The cosine modulation creates a characteristic DH PSF composed of two bright lobes, unequivocally defined by the respective fixed angles $\varphi$ and $\gamma$. The direction of zero-intensity axis in the resulting image is given by the angle $\theta = \pi/2 - \varphi - \gamma$. By setting the polarizer in the direction $\gamma = \pi/2$, the orientation of the lobes in the image plane provides an accurate estimate of the ground-truth angle of the nanorod in the object plane. In real conditions, a misalignment of the q-plate can slightly change the reference orientation of the DH PSF, resulting in a constant angular increment being added to the measured angles. This problem is easily solved by an initial one-time measurement on a calibration sample with known angular orientations of the dipole emitters, during which the q-plate misalignment can be eliminated by appropriately rotating the polarizer.

*Calibration measurement*

A proof-of-principle experiment was carried out in the optical system shown in Fig. 2. It was designed as a dark-field reflection microscope with an add-on module consisting of a q-plate and a linear polarizer. A dark-field illumination stop aperture (DIS) was projected into the back focal plane of a microscope objective (MO, Olympus 60×, NA 0.9) using an illumination lens (IL) so the sample was illuminated by plane waves with wave vectors forming a hollow cone. To excite the LSPR in nanorods of any orientation, a circular polarizer (CP) was inserted before the DIS. The measurement was carried out using quasi-monochromatic spatially incoherent light provided by a high-power LED (Thorlabs SOLIS-623C, 623 nm, full width at half maximum 50 nm). Back-reflected light was collected by the same MO, and through a beamsplitter (BS) directed towards a tube lens (TL$_1$). In the back focal plane of the TL$_1$, a magnified image of the sample was created and then transformed by the Fourier lens (FL). At the back focal plane of the FL, the spatial-frequency spectrum of the sample was accessible for filtering. By placing a dark-field stop aperture (DS) into this plane, any background light was suppressed and only the light scattered by nanorods was transmitted. This light then passed through a q-plate (Thorlabs WPV10L-633) and a linear polarizer (LP). Diffraction images of individual nanorods were subsequently created by a second tube lens (TL$_2$) and captured by a CMOS camera (Basler acA2040-90um).

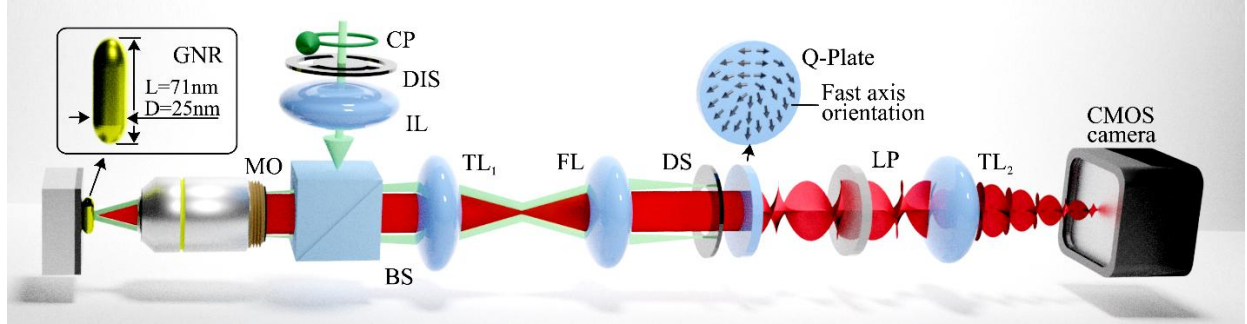

**Figure 2** Scheme of the experimental setup used for the orientation imaging: circular polarization CP, dark-field illumination stop aperture DIS, illumination lens IL, gold nanorod GNR, microscope objective MO, beam splitter BS, first tube lens TL$_1$, Fourier lens FL, dark-field stop aperture DS, SAM to OAM converter (Q-Plate), linear polarizer LP, second tube lens TL2 and complementary metal-oxide-semiconductor (CMOS) camera.

The orientation imaging procedure described above was experimentally tested using a benchmark sample composed of two rows of gold nanorods (length 200 nm, width 80 nm, height 30 nm), fabricated by electron beam lithography onto a silicon substrate.[42] Nanorods placed in the first row were gradually rotated by 22.5°, while all nanorods in the second row kept a constant orientation angle (see scanning electron microscope images in Fig. 3A). Note the scale bar in Fig. 3A shows the true size of the nanoantennas, but the separation distances do not match this scale (in-scale image is available in Supplementary material). The optical image of the benchmark sample provided by a conventional dark-field microscopy is presented in Fig. 3B. The symmetrical image spots confirm that the shape of individual nanorods is not resolved, and the diffraction-limited PSFs corresponding to point sources are obtained instead. The orientation imaging of the benchmark sample using the SAM to OAM conversion method proposed in this work is presented in Fig. 3C. Thanks to the known orientation of the nanorods, the add-on module could be calibrated using the linear polarizer (LP), so the angle of the DH PSF lobes coincided with the nanorod axes. To assess the measurement accuracy, the orientation of nanoantennas was reconstructed from the experimental data shown in Fig. 3C (angle $\varphi'$) and set against their ground-truth angular orientation ($\varphi$). The results are shown in Fig. 3D, where the inset quantifies deviations of the second-row reference DH PSFs with respect to the theoretical orientation. Using all 18 nanorods in Fig. 3C, the measurement accuracy (i.e., standard deviation from the ground-truth value) was estimated as 4.3°. It should be noted that some nanorods exhibit obvious shape irregularities that might adversely affect this accuracy value. The theoretically predicted robustness of our method to defocusing was also verified experimentally. In Fig. 3E we show an image of a single nanorod that was defocused in the range of ±3 μm. The defocusing caused only the blurring of the DH PSF, but its angular orientation remained unchanged.

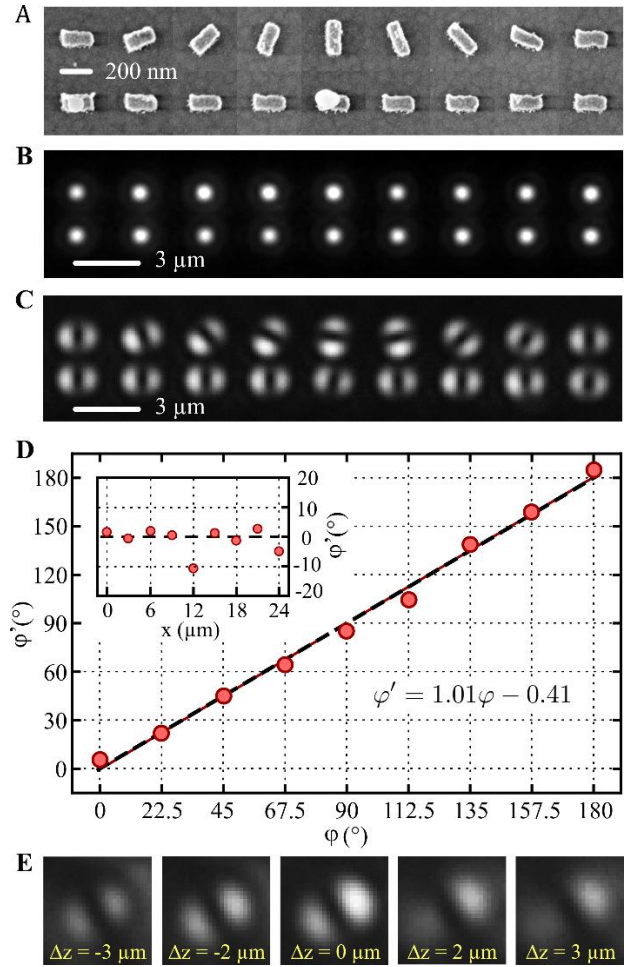

**Figure 3** Imaging of a benchmark sample and evaluation of the measurement accuracy. (A) Compound SEM image of gold nanorods used as a benchmark sample. The scale bar represents the true size of the nanorods, but their separation distances do not match this scale (the raw uncropped image is available in Supplementary material). (B) Standard dark-field image of the benchmark sample. (C) Image of the benchmark sample captured using the SAM to OAM conversion method proposed in this work. (D) Reconstruction of nanorod orientations $\varphi'$ from image (C) plotted against their theoretical orientation $\varphi$ (solid and dashed line correspond to theory and fit of the measured data, respectively). Inset shows deviations of reference DH PSFs from their theoretical values marked by a dashed line. (E) Images of a gold nanorod defocused in the range ±3 μm demonstrating that orientation of the DH PSF is insensitive to defocusing.

*Orientation imaging of sub-100 nm gold nanorods*

The real-world practical applicability of the proposed method was demonstrated by orientation imaging of colloidal gold nanorods randomly dispersed onto a sample. A commercial nanorod suspension (Nanopartz A12-25-650, diameter 25 nm, length 71 nm) was drop-casted onto a glass substrate covered by indium tin oxide, let to fully dry out, and subsequently imaged using our developed method. A typical image is presented in Fig. 4A. The measurement reliability and sensitivity were again verified using reference imaging in SEM. By comparing the orientation imaging of the nanorods shown in Fig. 4B with the reference SEM image of the same area presented in Fig. 4C, a good visual compliance is apparent. To assess these results quantitatively, an image-processing algorithm was developed (see Supplementary material), so the angular orientations of each DH PSF and the SEM nanorod image could be computationally reconstructed and compared. According to this comparison, the nanorod angles obtained by our orientation imaging method had accuracy (standard deviation) better than 2.5° (see Supplementary material). Situation deserving attention occurs when two nanorods assume an L-shaped configuration [area (d) in Fig. 4C]. Thanks to the LSPR of nanorods with the orthogonal orientation and to their sub-diffraction distance, the double-lobe spot is transformed to a ring-shaped vortex PSF [area (d) in Fig. 4B]. The ring-shaped vortex PSF is created thanks to incoherent superposition of two orthogonal DH PSFs, whose minimal lateral displacement is negligible in the optical image.

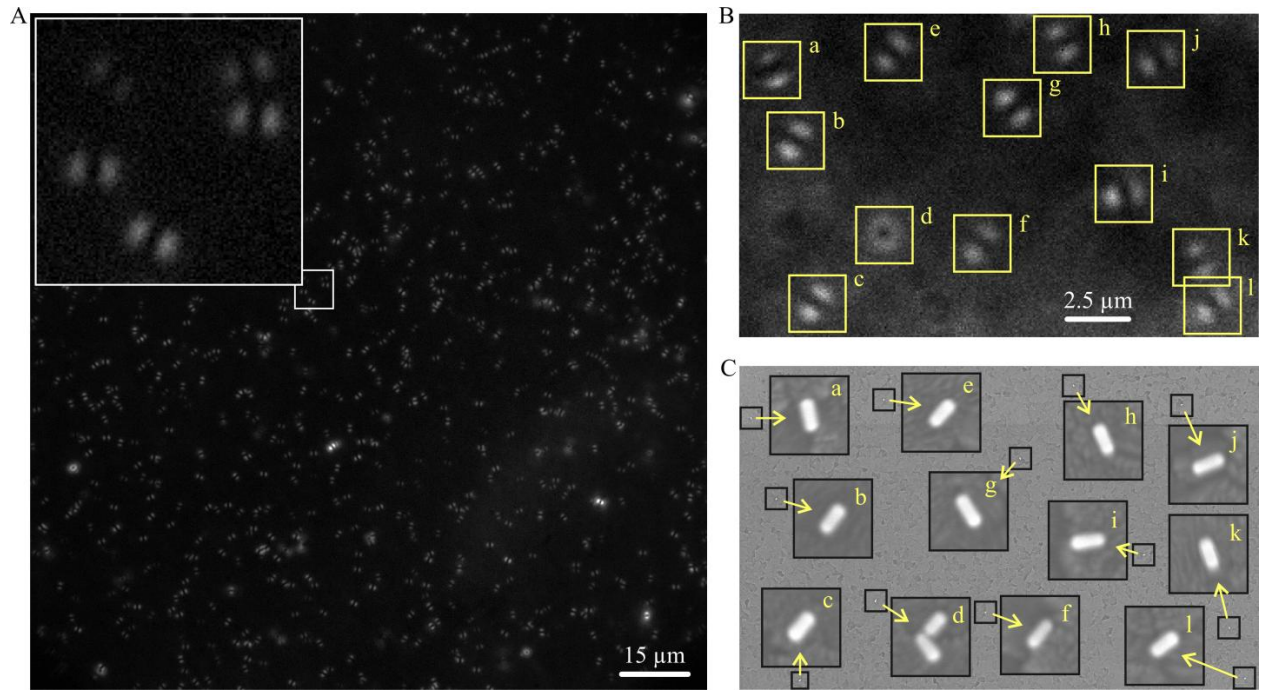

**Figure 4** Measurement of angular positions of randomly distributed sub-100 nm colloidal gold nanorods on ITO-coated glass substrate using the orientation imaging. (A) Full field of view image. (B) Close-up image for comparison with a reference SEM image. (C) Reference SEM image of the same field of view as in (B).

The ultimate test of the proposed method was a wide-field real-time orientation imaging of gold nanorods suspended in deionized water. The nanorods constantly moved by Brownian motion and the images were recorded with a rate of 7.7 frames per second. The frame rate was limited by illumination intensity and low CMOS sensor sensitivity, which could be both significantly improved by using a more powerful light source or an intensified CCD camera. The nanorods were measured in a large field of view (image of the full field of view is in Supplementary material), similar to the one shown in Fig. 4A. The rotational motion of nanorods in a selected part of the field of view is documented in Fig. 5 by a series of images taken at different times. In-plane rotation of a single gold nanorod is demonstrated in Fig. 5A using images taken with a time step of 1.3 seconds. The nanorod angle reconstructed from the DH PSF is indicated in the upper right corner of the individual images (Movie 1 with full temporal resolution is in Supplementary material). This experiment verifies the proposed possibility of monitoring in-plane rotational motion of a single nanorod by ambiguous-free real-time evaluation of its angular orientation. In

Fig. 5B, an image of a small cluster of nanorods is shown. The nanorods move collectively and their images are well distinguished even in near positions thanks to the spatially incoherent illumination suppressing the interference between the created DH PSFs. Images in Fig. 5B are separated by a time step of 1.3 seconds (Movie 2 with full temporal resolution is in Supplementary material). In Fig. 5C, a special type of motion of a single nanorod is displayed, in which the double-lobe DH PSF not only rotates but also morphs into a ring-shaped PSF at specific times, as marked by the yellow arrows (Movie 3 is in Supplementary material). We are convinced that the ring-shaped vortex PSF originates from the out-of-plane nanorod rotation, in which the dominance of the LSPR is lost and the influence of the transverse SPR increases. Note the same situation would occur for nanorods with smaller aspect ratio and therefore higher symmetry. In that case, the scattered intensity coming from the transverse SPR and LSPR has a comparable strength, two perpendicular DH PSFs are incoherently superposed and the ring-shaped PSF is created. The ring-shaped spot [area (d) in Fig. 4B] generated by nanorods in L-shaped configuration [area (d) in Fig. 4C] provides experimental evidence for this argument. One can therefore envision that estimation of nanorod aspect ratios or evaluation of their out-of-plane rotation will be possible via processing azimuthal intensity variations of the DH PSF. Thorough experimental validation of these effects is beyond the scope of this Letter and a comprehensive theoretical model with validating experiments will be presented elsewhere.

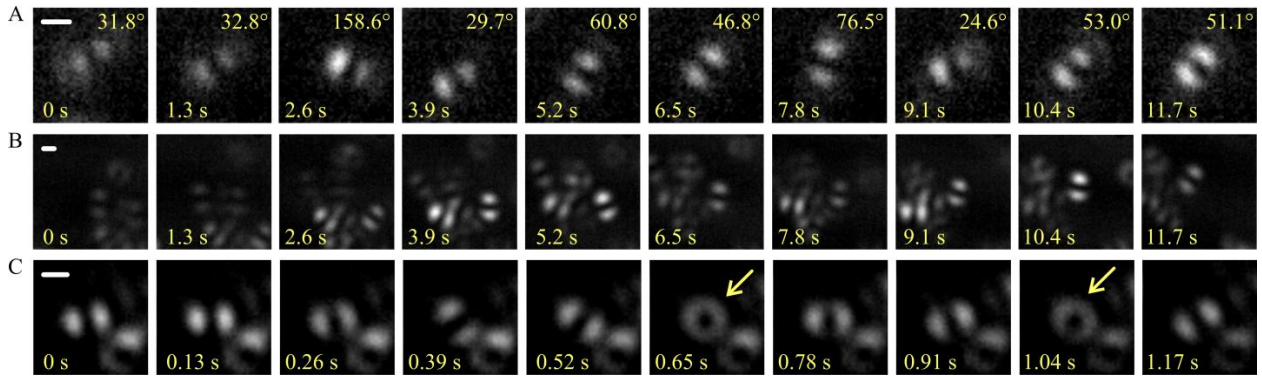

**Figure 5** Orientation measurement of sub-100 nm gold nanorods (diameter 25 nm, length 71 nm) suspended in deionized water (scale bars correspond to 1 μm). (A) In-plane rotation of a single nanorod documented by images taken with a time step of 1.3 seconds. Angular orientations reconstructed using an image-processing algorithm are indicated in the top-right corners. (B) Collective movement of a cluster of nanorods, which can be separated into the motion of individual nanorods thanks to incoherent illumination. (C) Rotation of a DH PSF and its transformation to a ring-shaped vortex PSF (marked by arrows), caused by out-of-plane rotation of the gold nanorod.

**Conclusions**

In conclusion, we have demonstrated a new optical technique allowing a single-shot wide-field orientation imaging of elongated nanoparticles made of noble metals. Our method is based on manipulations with the angular momentum of light, namely on SAM to OAM conversion by a geometric-phase element (q-plate). The accuracy of our method was evaluated using a lithographically fabricated benchmark sample with known orientations of gold nanorods, and the standard deviation fell below 4.5°. The practical applicability of the method was demonstrated by orientation imaging of randomly deposited sub-100 nm colloidal gold nanorods (diameter 25 nm, length 71 nm) on a solid substrate and by real-time monitoring of nanorods suspended in deionized water. During orientation imaging of colloidal nanorods, the obtained results were compared with ground-truth values provided by SEM, from which accuracy (standard deviation) 2.5° was achieved. Experiments proved the capability to real-time tracking of rotational movement while preserving high measurement accuracy, the parameters critical for study of dynamic phenomena occurring in practical applications. The SAM to OAM orientation imaging can be extended towards out-of-plane orientations and also to nanorod aspect ratio estimation. As it is not limited to the LSPR light, an extension towards characterization of any dipolar emitters of sub-diffraction size is also possible.

## Supporting Information Available

Scanning electron microscopy images of gold nanorods and optical images of gold nanorods acquired using the proposed method, workflow of the numerical reconstruction of nanorod orientation from optical images, and demonstration of the real-time optical imaging of gold nanorods (PDF, MPG).

## Acknowledgement

This work has been supported by the Grant agency of the Czech Republic (18-01396S, 20-01673S), Ministry of Education, Youth and Sports of the Czech Republic (LQ1601, CEITEC 2020), Palacký University (IGA PrF 2020 004). We also acknowledge CzechNanoLab Research Infrastructure supported by MEYS CR (LM2018110). PB has been supported by a scholarship awarded by the Czechoslovak Microscopy Society.

## Author's contributions

P.B and Z.B. conceived the idea. T.F., M.B., and P.B. performed optical measurements. P.S. was responsible for data processing and reconstruction of nanorods orientation. P.D., K.R., and F.L. prepared samples for experiments and performed reference SEM measurements. P.B. and Z.B wrote the manuscript. Z.B. supervised the Palacký University team. T.Š. and R.Ch. supervised the Brno University of Technology team. PB coordinated activities of participating laboratories. All authors discussed the results and commented on the manuscript.

## Disclosures

The authors declare no conflicts of interest.

Supplementary material

# Single-Shot Orientation Imaging of Nanorods Using Spin-to-Orbital Angular Momentum Conversion of Light

Tomáš Fordey,[1] Petr Bouchal,[2,3,*] Michal Baránek,[1] Petr Schovánek,[1] Zdeněk Bouchal,[1] Petr Dvořák,[2,3] Katarína Rovenská,[2,3] Filip Ligmajer,[2,3] Radim Chmelík,[2,3] Tomáš Šikola[2,3]

[1]Department of Optics, Palacký University, 17. listopadu 1192/12, 771 46 Olomouc, Czech Republic
[2]Institute of Physical Engineering, Faculty of Mechanical Engineering, Brno University of Technology, Technická 2, 616 69 Brno, Czech Republic
[3]Central European Institute of Technology, Brno University of Technology, Purkyňova 656/123, 612 00 Brno, Czech Republic

*petr.bouchal@ceitec.vutbr.cz

*Calibration measurement*

Proof of the principle experiment was realized using a benchmark sample prepared lithographically as an array of gold nanorods with gradually changing and fixed angular orientation. In the main text, a magnified detail of the individual nanorods was demonstrated for better clarity, in which their separation distance does not match the scale used. In Fig. S1A, the scanning electron microscopy (SEM) image of the benchmark sample is shown at a scale unified for both nanorod size and spacing. To maintain a high resolution in a large field of view, the image was divided into 7 subfields, which were moved together to form Fig. 3A in the main text. Details of individual nanoantennas are shown in Fig. S1B.

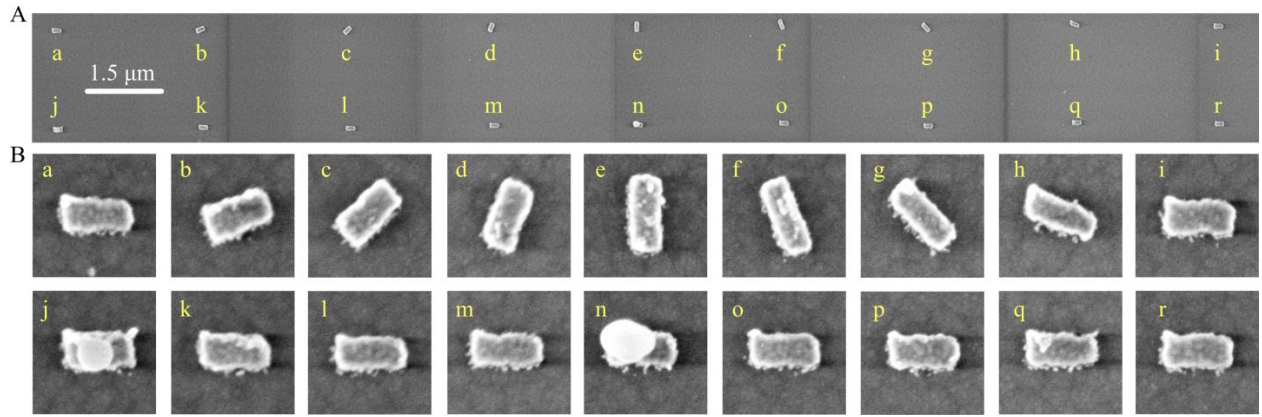

**Figure S1** In-scale image of benchmark sample used for calibration of the orientation imaging. (A) The whole area of the benchmark sample. (B) Details of individual gold nanorods.

*Orientation imaging of colloidal gold nanorods on Indium Tin Oxide (ITO) surface*

Orientation imaging of colloidal gold nanorods was realized using indium tin oxide (ITO) coated glass as a substrate. The optical images captured using ITO coated glass substrates were slightly affected by adverse noise effects due to the structure of the ITO surface. This structure is clearly visible in Fig. S2A,B, showing SEM images of colloidal gold nanorods placed on the ITO surface. Despite the background light, orientation imaging and numerical reconstruction of nanorod images were possible.

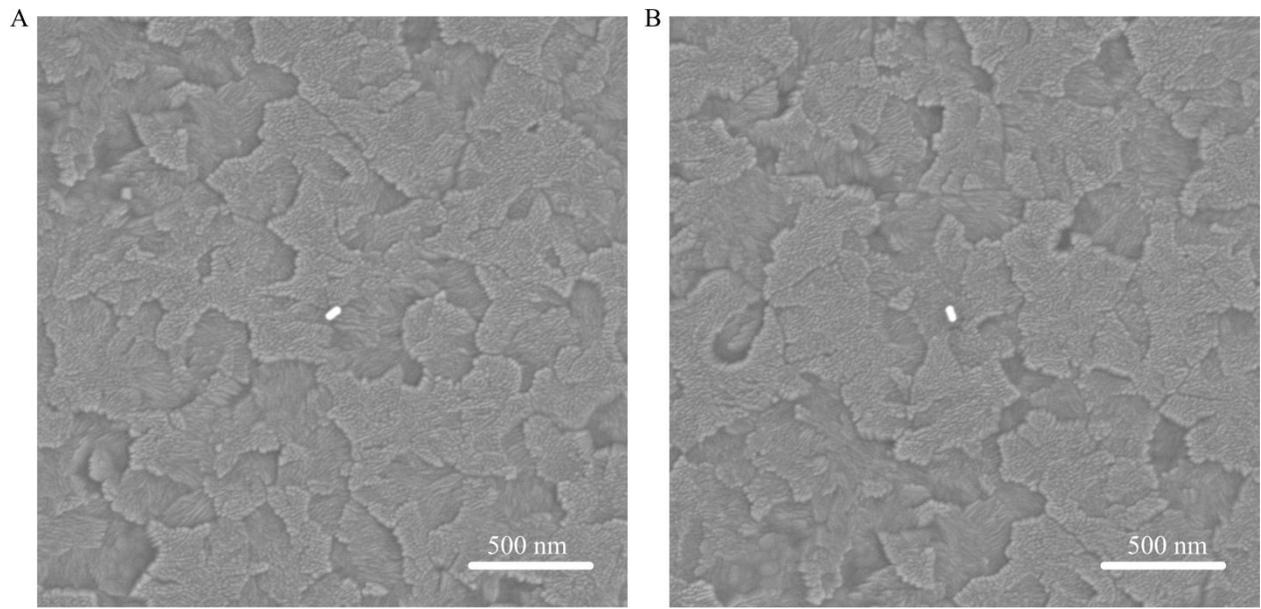

**Figure S2** SEM images of gold nanorods on ITO coated glass substrate showing the structure of the ITO surface.

*Numerical reconstruction of DH PSF images*

In the proposed SAM to OAM conversion method, the orientation of nanorods is determined from a spatially structured diffraction image spot created by the interference of optical vortices. The image corresponds to a double-helix point spread function (DH PSF) formed by two lobes with the angular position determining the nanorod orientation. The orientation of the nanorod thus can be reconstructed by numerical processing of measured DH PSFs. The measurement reliability and sensitivity were verified using reference imaging in SEM. Using image processing in Mathematica software, the orientation of the DH-PSF was restored and then compared to the nanorod orientation measured by SEM.

The measured optical image was pre-processed using a median filter 2-pixel range neighborhood. This allows us to smooth the image and reduce the noise. Using adaptive thresholding, the binary mask was created, allowing segmentation and labeling of individual DH PSFs (a-l in Fig. S3A). After that, the raw image was multiplied by the binary mask. The result is shown in Fig. S3B. The labeled images of individual DH PSFs were gradually processed and their orientation was reconstructed. During the reconstruction process, the positions of intensity centroids of DH PSF lobes were calculated ($T_1$ and $T_2$ in Fig. S3C) and the orientation of the vector x' (pointing from $T_1$ to $T_2$) with respect to the x-axis of the Cartesian coordinate system was evaluated.

A similar processing was applied to the SEM image. Instead of calculating centroid from segmented and labeled images of nanorods, the axes of symmetry x' and y' were calculated (Fig. S4) determining the nanorod orientation.

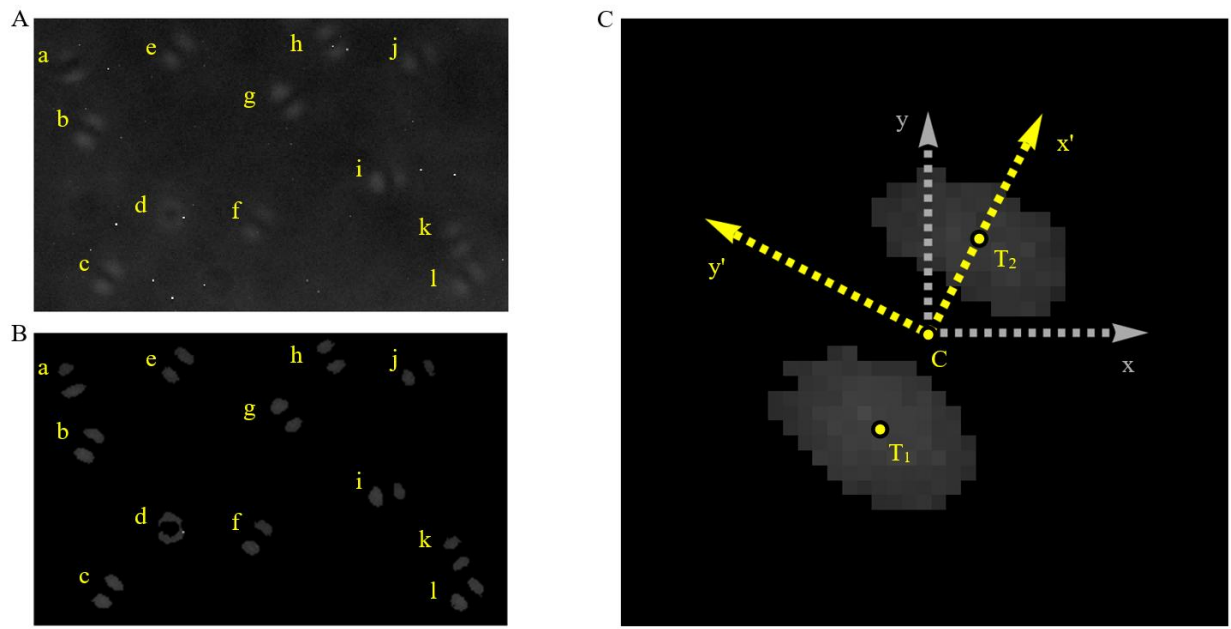

**Figure S3** The workflow of numerical reconstruction of nanorod orientation from DH PSF images. (A) Raw optical image of colloidal gold nanorods captured using the proposed method. (B) Separation of DH PSFs from the background using adaptive thresholding and binary mask. (C) Reconstruction of DH PSF orientation [marked b in (A) and (B)] using intensity centroids T1 and T2 of DH PSF lobes.

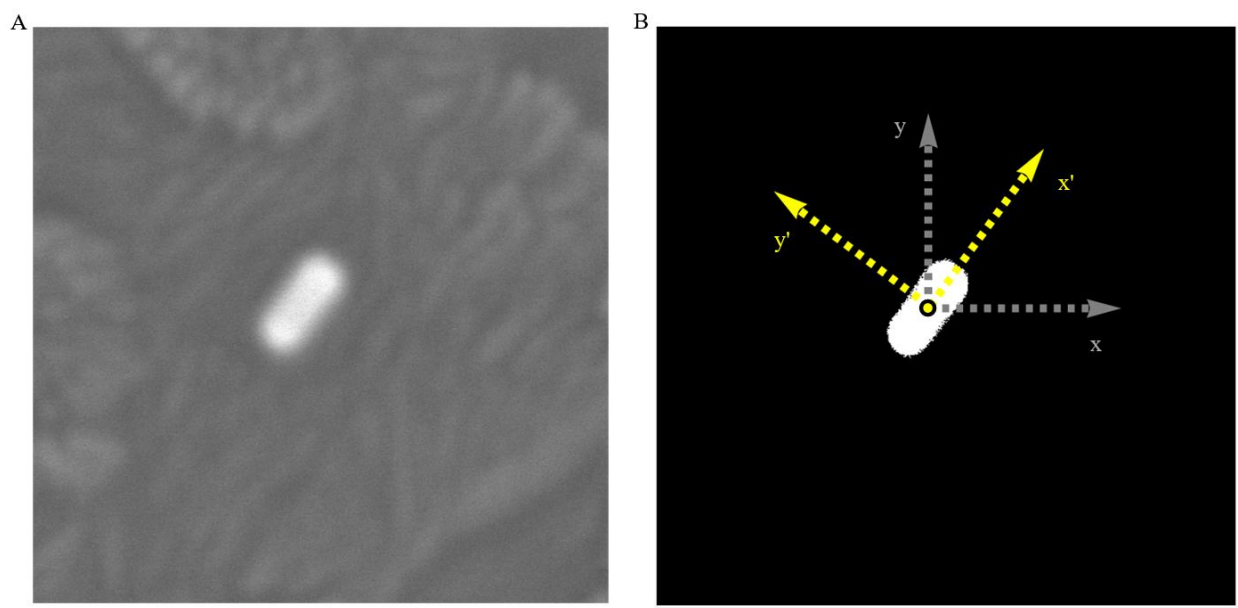

**Figure S4** Reconstruction of nanorod orientation from SEM image. (A) SEM image of colloidal gold nanorod on the ITO surface. (B) Binarized image and determined axes of symmetry x' and y' of the nanorod in the reference coordinate system.

The described procedures were applied to the images provided by the proposed SAM to OAM conversion method and SEM (Fig. 4B,C in the main text), and the orientation angles of nanorods $\varphi'_{DHPSF}$ and $\varphi'_{SEM}$ were determined. The angles $\varphi'_{SEM}$ were adopted as the ground-truth values and the standard deviation of 2.5° was calculated from the differences $\Delta\varphi' = \varphi'_{DH\ PSF} - \varphi'_{SEM}$. The angles $\varphi'_{SEM}$, $\varphi'_{DHPSF}$ and their differences $\Delta\varphi'$ are given in Table 1 for all measured nanoantennas denoted as (a)-(l) in Fig. 4B,C of the main text. The results show that a systematic difference with the mean value of 5.5° is introduced between the angular nanorod position and the orientation angle of the DH PSF. This constant bias does not affect the measurement accuracy and can be eliminated by adjusting the polarizer if necessary.

|  | a | b | c | d | e | f | g | h | i | j | k | l |
|---|---|---|---|---|---|---|---|---|---|---|---|---|
| φ'$_{SEM}$ (°) | 101.6 | 53.8 | 51.2 | X | 48.0 | 52.1 | 121.1 | 112.9 | 8.0 | 22.3 | 108.6 | 38.0 |
| φ'$_{DH\ PSF}$ (°) | 111.2 | 59.5 | 57.9 | X | 55.9 | 55.7 | 130.0 | 117.0 | 11.3 | 23.9 | 112.0 | 44.0 |
| Δφ' (°) | 9.6 | 5.7 | 6.7 | X | 7.9 | 3.6 | 8.9 | 4.1 | 3.3 | 1.6 | 3.4 | 6.0 |

Table 1 Reconstruction of the nanorod orientation using the SEM and rotation imaging.

*Orientation imaging of colloidal gold nanorods suspended in deionized water*

In the main text, the real-time orientation imaging of colloidal gold nanorods was demonstrated using selected parts of the field of view and images taken at different times (Fig. 5 in the main text). Fig. S5 shows one representative frame of the full field of view from which the Supplementary movies were taken.

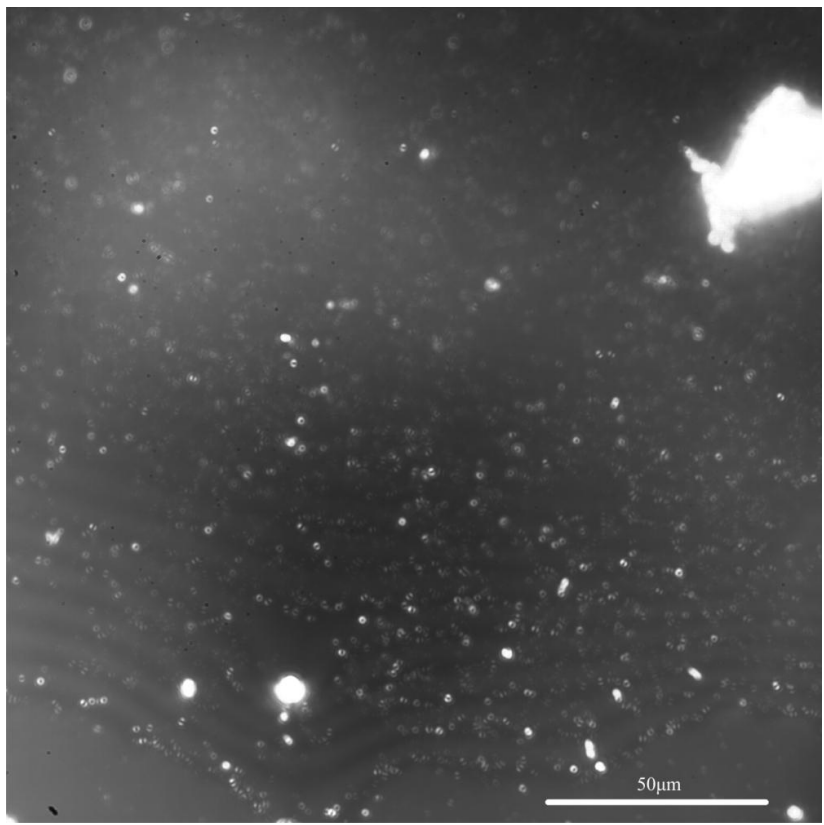

Figure S5 Representative image showing the size of the field of view recorded in the real-time orientation imaging of colloidal gold nanorods suspended in deionized water.